\documentclass[groupedaddress,aps,pra,superscriptaddress,showpacs,twocolumn,prl]{revtex4}%
\usepackage{epsfig,dsfont,amssymb,amsmath,amsthm,amsfonts,amsbsy,mathrsfs}
\usepackage{graphicx, color}
\usepackage{epstopdf}
\usepackage{mathdots}
\usepackage{float}
\begin{document}

\title{Tighter generalized monogamy and polygamy relations for multiqubit systems}

\author{Zhi-Xiang Jin}
\thanks{Corresponding author: jzxjinzhixiang@126.com}
\affiliation{School of Mathematical Sciences,  Capital Normal University,  Beijing 100048,  China}
\author{Shao-Ming Fei}
\thanks{Corresponding author: feishm@mail.cnu.edu.cn}
\affiliation{School of Mathematical Sciences,  Capital Normal University,  Beijing 100048,  China}
\affiliation{Max-Planck-Institute for Mathematics in the Sciences, Leipzig 04103, Germany}

\bigskip

\begin{abstract}

We present a different kind of monogamy and polygamy relations based on concurrence and concurrence of assistance for multiqubit systems. By relabeling the subsystems associated with different weights, a smaller upper bound of the $\alpha$th ($0\leq\alpha\leq2$) power of concurrence for multiqubit states is obtained. We also present tighter monogamy relations satisfied by the $\alpha$th ($0\leq\alpha\leq2$) power of concurrence for $N$-qubit pure states under the partition $AB$ and $C_1 . . . C_{N-2}$, as well as under the partition $ABC_1$ and $C_2\cdots C_{N-2}$. These inequalities give rise to the restrictions on entanglement distribution and the trade off of entanglement among the subsystems. Similar results are also derived for negativity.

\end{abstract}

\maketitle

\section{INTRODUCTION}

Quantum entanglement \cite{FMA,KSS,HPB,HPBB,JIV,CYS} is an essential feature of quantum mechanics, which distinguishes the quantum from the classical world. One of the fundamental differences between classical and quantum correlations lies on the sharability among the subsystems. Different from the classical correlation, quantum correlation cannot be freely shared. The monogamy relations give rise to the restrictions on the distribution of entanglement in the multipartite setting. It is not possible to prepare three qubits in a way that any two qubits are maximally entangled.
The monogamy relation was first quantified by Coffman, Kundu, and Wootters (CKW) \cite{MK} for three qubits, $\mathcal{E}_{A|BC}\geq \mathcal{E}_{AB} +\mathcal{E}_{AC}$, where $\mathcal{E}_{A|BC}$ denotes the entanglement between systems $A$ and $BC$.
The CKW inequality shows that the more entanglement shared between two qubits $A$ and $B$, the less entanglement between the qubits $A$ and $C$. CKW inequality was generalized to multiqubit systems \cite{TJ,ZXN,JZX,jll} and also studied intensively in more general settings \cite{gy1,gy2}.

 Using concurrence of assistance \cite{tfs} as the measure of distributed entanglement, the polygamy of entanglement
provides a lower bound for the distribution of bipartite entanglement in a multipartite system \cite{bcs}.
Polygamy of entanglement is characterized by the polygamy inequality, ${E_a}_{A|BC}\leq {E_a}_{AB} +{E_a}_{AC}$ for a tripartite quantum state $\rho_{ABC}$, where ${E_a}_{A|BC}$ is the assisted entanglement \cite{gg}
between $A$ and $BC$. Polygamy of entanglement was generalized to multiqubit systems \cite{bcs} and arbitrary dimensional multipartite states \cite{062328,295303,bcs,042332}. In Ref. \cite{zhu}, the authors have given the monogamy and polygamy relations with any qubits as the focus ones for multiqubit states. Furthermore, the case of the $\alpha$th ($0\leq\alpha\leq2$) power of concurrence for $N$-qubit pure states under any partition was studied in \cite{jin}.

In this paper, we study the general monogamy inequalities with qubits $AB$ as the focus qubits, satisfied by the concurrence and the concurrence of assistance (COA).
A smaller (tighter) upper bound for the $\alpha$th ($0\leq\alpha\leq2$) power of concurrence for multiqubit states is obtained. Then we establish the tighter monogamy relations of the $\alpha$th ($0\leq\alpha\leq2$) power of concurrence in $N$-qubit pure states under the partition $AB$ and $C_1 . . . C_{N-2}$, as well as under the partition $ABC_1$ and $C_2\cdots C_{N-2}$. Based on the relations between negativity and concurrence, we also obtain similar results for negativity.  Detailed examples are presented.

\section{Tighter generalized monogamy and polygamy relations of concurrence}
Let $H_X$ denote the finite dimensional vector space associated with qubit $X$.
For a bipartite pure state $|\psi\rangle_{AB}$ in vector space $H_A\otimes H_B$, the concurrence is given by \cite{AU,PR,SA}
\begin{equation}\label{CD}
C(|\psi\rangle_{AB})=\sqrt{{2\left[1-\mathrm{Tr}(\rho_A^2)\right]}},
\end{equation}
where $\rho_A$ is the reduced density matrix by tracing over the subsystem $B$, $\rho_A=\mathrm{Tr}_B(|\psi\rangle_{AB}\langle\psi|)$. The concurrence for a bipartite mixed state $\rho_{AB}$ is defined by the convex roof
\begin{equation*}
 C(\rho_{AB})=\min_{\{p_i,|\psi_i\rangle\}}\sum_ip_iC(|\psi_i\rangle),
\end{equation*}
where the minimum is taken over all possible decompositions of $\rho_{AB}=\sum_ip_i|\psi_i\rangle\langle\psi_i|$, with $p_i\geq0$, $\sum_ip_i=1$ and $|\psi_i\rangle\in H_A\otimes H_B$.

For a tripartite state $|\psi\rangle_{ABC}$, the concurrence of assistance (COA) is defined by \cite{CH}
\begin{equation*}
 C_a(|\psi\rangle_{ABC})=C_a(\rho_{AB})=\max_{\{p_i,|\psi_i\rangle\}}\sum_ip_iC(|\psi_i\rangle),
\end{equation*}
for all possible ensemble realizations of $\rho_{AB}=\mathrm{Tr}_C(|\psi\rangle_{ABC}\langle \psi|)=\sum_ip_i|\psi_i\rangle\langle\psi_i|$. When $\rho_{AB}$ is a pure state, one has $C(|\psi\rangle_{AB})=C_a(\rho_{AB})$.

For an $N$-qubit state $|\psi\rangle_{AB_1,\cdots,B_{N-1}}\in H_A\otimes H_{B_1}\otimes\cdots\otimes H_{B_{N-1}}$, the concurrence $C(|\psi\rangle_{A|B_1\cdots B_{N-1}})$ of the state $|\psi\rangle_{A|B_1\cdots B_{N-1}}$, viewed as a bipartite partition $A$ and $B_1B_2\cdots B_{N-1}$, satisfies the monogamy inequality \cite{TF},
\begin{eqnarray}\label{CAA}
 &&C^{2}(\rho_{A|B_1,B_2\cdots,B_{N-1}})\nonumber\\&&\geq C^{2}(\rho_{AB_1})+C^{2}(\rho_{AB_2})+\cdots+C^{2}(\rho_{AB_{N-1}}),
\end{eqnarray}
where $C(\rho_{AB_i})$ is the concurrence of $\rho_{AB_i}=\mathrm{Tr}_{B_1\cdots B_{i-1}B_{i+1}\cdots B_{N-1}}(\rho)$.

The dual inequality satisfied by COA for $N$-qubit states has the form \cite{GSB},
\begin{eqnarray}\label{DCA}
 C^2(|\psi\rangle_{A|B_1B_2\cdots B_{N-1}}) \leq \sum_{i=1}^{N-1}C_a^2(\rho_{AB_i}).
\end{eqnarray}

Furthermore, the authors in \cite{ZXN} presented a generalized monogamy relation for $\alpha\geq2$,
$C^{\alpha}(\rho_{A|B_1,B_2\cdots,B_{N-1}})\geq C^{\alpha}(\rho_{AB_1})+C^{\alpha}(\rho_{AB_2})+\cdots+C^{\alpha}(\rho_{AB_{N-1}})$. The dual inequality is given in \cite{jin} for $0\leq \alpha \leq 2$,
\begin{eqnarray}\label{jzx}
&&C^\alpha(|\psi\rangle_{A|B_1B_2\cdots B_{N-1}})\leq \nonumber\\&&C^\alpha_a(\rho_{AB_1})+\frac{\alpha}{2}C_a^\alpha(\rho_{AB_2})+\cdots\nonumber\\&&
 +\left(\frac{\alpha}{2}\right)^{N-2}C_a^\alpha(\rho_{AB_{N-1}}).
 \end{eqnarray}
In this paper, we first give a tighter upper bound satisfied by the $\alpha$th power of COA for $N$-qubit states. Then we prsent monogamy and polygamy relations for $N$-qubit states in terms of the $\alpha$th power of COA, which are tighter than the existing ones.

The concurrence (\ref{CD}) is related to the linear entropy $T(\rho)$ of a state $\rho$, $T(\rho)=1-\mathrm{Tr}(\rho^2)$ \cite{EM}.
For a bipartite state $\rho_{AB}$, $T (\rho)$ has the property \cite{CYY},
\begin{eqnarray}\label{LS}
|T(\rho_A)-T(\rho_B)|\leq T(\rho_{AB})\leq T(\rho_A)+T(\rho_B).
\end{eqnarray}
For convenience, we rewrite (\ref{DCA}) as follows,
\begin{eqnarray}\label{coa}
 C^2(|\psi\rangle_{A|B_1B_2\cdots B_{N-1}}) \leq \sum_{i=1}^{k}C_a^2(\rho_{AM_i}),
\end{eqnarray}
where $C_a^2(\rho_{AM_i})=\sum_{j=M_{i-1}+1}^{M_i}C_a^2(\rho_{AB_j})$ with $M_0=0,~\sum_{i=1}^kM_i=N-1$, $1\leq k\leq N-1$.
The summation on the right hand side of (\ref{coa}) has been separated into $k$ parts.  There is always a choice of $M_i$, such that the above relations is true. 

{\bf[Theorem 1]}. For any $N$-qubit pure state $|\psi\rangle_{AB_1B_2\cdots B_{N-1}}$, we have
\begin{eqnarray}\label{le}
  &&C^\alpha(|\psi\rangle_{A|B_1B_2\cdots B_{N-1}})\nonumber\\[1mm]
 &&\leq  C^\alpha_a(\rho_{AM_1})+h C_a^\alpha(\rho_{AM_2})+\cdots\nonumber\\[1mm]
 &&
  +h^{k-1}C_a^\alpha(\rho_{AM_k}),
\end{eqnarray}
for all $0\leq \alpha\leq2$, where $h=2^\frac{\alpha}{2}-1$.

{\sf[Proof]}. Without loss of generality, we can always assume that $C_a^2(\rho_{AM_t})\geq \sum_{l=t+1}^k C_a^2(\rho_{AM_l})$, $1\leq t\leq k-1,~2\leq k\leq N-1$,  by reordering
$M_1,M_2,\cdots,M_k$ and/or relabeling the subsystems in need.
Form the result in \cite{GSB}, we have
\begin{eqnarray}\label{pfth1}
 &&C^\alpha(|\psi\rangle_{A|B_1B_2\cdots B_{N-1}}) \nonumber\\
 &&\leq \left(C_a^2(\rho_{AM_1})+\sum_{i=2}^{k}C_a^2(\rho_{AM_i})\right)^{\frac{\alpha}{2}}\nonumber\\
 &&=C_a^\alpha(\rho_{AM_1})\left(1+\frac{\sum_{i=2}^{k}C_a^2(\rho_{AM_i})}{C_a^2(\rho_{AM_1})}\right)^{\frac{\alpha}{2}}\nonumber \\
   && \leq C_a^\alpha(\rho_{AM_1})\left[1+h\left(\frac{\sum_{i=2}^{k}C_a^2(\rho_{AM_i})}{C_a^2(\rho_{AM_1})}\right)^{\frac{\alpha}{2}}\right]\nonumber\\
   &&=C^\alpha_a(\rho_{AM_1})+h\left(\sum_{i=2}^{k}C_a^2(\rho_{AM_i})\right)^\frac{\alpha}{2}\nonumber\\
   &&\leq \cdots \leq\sum_{i=1}^kh^{i-1}C^\alpha_a(\rho_{AM_i}),
\end{eqnarray}
where the first inequality is due to (\ref{coa}). 
By using the fact that \cite{jzxf}, for any real numbers $x$ and $t$ such that $0\leq t \leq 1$ and $0\leq x \leq 1$, $(1+t)^x\leq 1+(2^{x}-1)t^x$, we get the second inequality. $\Box$

Theorem 1 gives a tighter polygamy relation of the $\alpha$th ($0\leq \alpha\leq2$) power of concurrence for $N$-qubit pure state $|\psi\rangle_{A|B_1B_2,\cdots,B_{N-1}}$ based on the COA. For the case of $k=N-1$, we have the following result,
\begin{eqnarray}\label{pfth1}
 &&C^\alpha(|\psi\rangle_{A|B_1B_2\cdots B_{N-1}})\nonumber\\[1mm]&&
 \leq C^\alpha_a(\rho_{AB_1})+hC_a^\alpha(\rho_{AB_2})+\cdots\nonumber\\[1mm]&&
 +h^{N-2}C_a^\alpha(\rho_{AB_{N-1}}).
\end{eqnarray}
Specially, for $\alpha=2$, inequality (\ref{le}) or (\ref{pfth1}) reduces to the result (\ref{coa}) in \cite{GSB}. For $0<\alpha <2$, inequality (\ref{le}) or (\ref{pfth1}) is tighter than the result (\ref{jzx}) in \cite{jin}.

{\it Example 1}. Let us consider the three-qubit state $|\psi\rangle$ in the generalized Schmidt decomposition form \cite{AA,XH},
\begin{equation*}
|\psi\rangle=\lambda_0|000\rangle+\lambda_1e^{i{\varphi}}|100\rangle+\lambda_2|101\rangle
+\lambda_3|110\rangle+\lambda_4|111\rangle,
\end{equation*}
where $\lambda_i\geq0,~i=0,\cdots,4$ and $\sum_{i=0}^4\lambda_i^2=1.$
We have $C(\rho_{A|BC})=2\lambda_0\sqrt{{\lambda_2^2+\lambda_3^2+\lambda_4^2}}$, $C(\rho_{AB})=2\lambda_0\lambda_2$, $C(\rho_{AC})=2\lambda_0\lambda_3$,  $C_a(\rho_{AB})=2\lambda_0\sqrt{{\lambda_2^2+\lambda_4^2}}$,  $C_a(\rho_{AC})=2\lambda_0\sqrt{{\lambda_3^2+\lambda_4^2}}$.
Set $\lambda_0=\lambda_1=\lambda_2=\lambda_3=\lambda_4=\frac{\sqrt{5}}{5}$. One gets $C^{\alpha}(\rho_{A|BC})=(\frac{2\sqrt{3}}{5})^{\alpha}$, $C_a^{\alpha}(\rho_{AB})+\frac{\alpha}{2}C_a^{\alpha}(\rho_{AC})=\left(1+\frac{\alpha}{2}\right)(\frac{2\sqrt{2}}{5})^{\alpha}$, $C_a^{\alpha}(\rho_{AB})+hC_a^{\alpha}(\rho_{AC})=2^\frac{\alpha}{2}(\frac{2\sqrt{2}}{5})^{\alpha}$. One can see that our result is better than that of (\ref{jzx}) for $0<\alpha <2$, see Fig. 1.
\begin{figure}
  \centering
  \includegraphics[width=8.5cm]{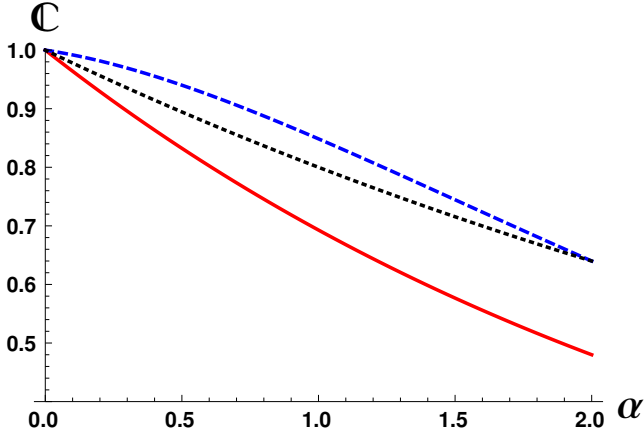}\\
  \caption{$E$ is the entanglement as a function of $\alpha$. Solid (Red) line is the $\alpha$th power of concurrence; Dotted (Black) line is the upper bound in (\ref{le}); Dashed (Blue) line is the result of (\ref{jzx}).}\label{1}
\end{figure}

In the following, by using the conclusion of Theorem 1 and Lemma 2, we present some monogamy-type inequalities and lower bounds of concurrence in terms of concurrence and COA. These monogamy relations are satisfied by the concurrence of $N$-qubit states under the partition $AB$ and $C_1\cdots C_{N-2}$, as well as under the partition $ABC_1$ and $C_2\cdots C_{N-2}$, which generalize the monogamy inequalities for pure states in \cite{ZXN} and give tighter bounds than \cite{jin}.

{\bf [Lemma]}. For arbitrary two real numbers $x$ and $y$ such that $x\geq y\geq0$, we have
 $(x-y)^\alpha\geq x^\alpha-y^\alpha$ and $(x+y)^\alpha\leq x^\alpha+y^\alpha$ for $0\leq\alpha\leq 1$.

{\sf [Proof]}. $(x-y)^\alpha\geq x^\alpha-y^\alpha$ is equivalent to $(1-\frac{y}{x})^\alpha+(\frac{y}{x})^\alpha\geq1$ for nonzero $x$. Denote $t=\frac{y}{x}$. Then $0\leq t\leq 1$. Set $f(t)=(1-t)^\alpha+t^\alpha$.
We have $\frac{\mathrm{d}f}{\mathrm{d}t}=\alpha[t^{\alpha-1}-(1-t)^{\alpha-1}]$.
For $0\leq t \leq \frac{1}{2}$, $\frac{\mathrm{d}f}{\mathrm{d}t}\geq 0$, since $1-t\geq t$ and $\alpha-1<0$. Therefore, $f(t)\geq f(0)=1$ in this case. For $\frac{1}{2}< t \leq 1$, $\frac{\mathrm{d}f}{\mathrm{d}t}\leq 0$, since $t\geq 1-t$ and $\alpha-1\leq 0$. Hence, $f(t)\geq f(1)=1$ in this case.
In summary, for $0<\alpha\leq 1$, $f(t)_\mathrm{min}\geq f(0)=f(1)=1$. Similarly, one can get the second inequality in Lemma. When $\alpha=0$ or $x=0$, the inequality is trivial. Hence we complete the proof of the Lemma. $\Box$

{\bf [Theorem 2]}. For any $N$-qubit state $|\psi\rangle_{ABC_1\cdots C_{N-2}}$, we have
\begin{eqnarray}\label{thm2}\nonumber
&&C^\alpha(\rho_{AB|C_1\cdots C_{N-2}})\\ \nonumber
&&\geq\mathrm{max}\left\{h\sum_{i=1}^{k_1-1}C^\alpha(\rho_{AM_i})+C^\alpha(\rho_{AM_{k_1}})-J_B,\right.\\
&& \left. h\sum_{i=1}^{k_2-1}C^\alpha(\rho_{BM_i})+C^\alpha(\rho_{BM_{k_2}})-J_A\right\},
\end{eqnarray}
for $0\leq\alpha\leq2$, $N\geq 4$,
where $C_a^2(\rho_{AM_i})$ is defined in (\ref{coa}) and  $J_A= \sum_{i=1}^{k_1}h^{i-1}C^\alpha_a(\rho_{AM_i})$, $J_B=\sum_{i=1}^{k_2}h^{i-1}C^\alpha_a(\rho_{BM_i})$, $h=2^\frac{\alpha}{2}-1$, $k_1,k_2$ are defined similar to inequality (\ref{coa}).

{\sf [Proof]}. Without loss of generality, there always exists a proper ordering of the subsystems $M_{t_i},M_{t_i+1},\cdots,M_{k_i}$ $(i=1,2)$ such that $C_a^2(\rho_{AM_{t_1}})\geq \sum_{l={t_1}+1}^{k_1} C_a^2(\rho_{AM_l})$ and $C_a^2(\rho_{BM_{t_2}})\geq \sum_{l={t_2}+1}^{k_2} C_a^2(\rho_{BM_l})$, $1\leq t_i\leq k_i-1$ $(i=1,2)$, $2\leq k_1,k_2\leq N-1$.

For $N$-qubit pure state $\rho_{ABC_1\cdots C_{N-2}}$, if $C(\rho_{A|BC_1\cdots C_{N-2}})\geq C(\rho_{B|AC_1\cdots C_{N-2}})$, one has
\begin{eqnarray*}
&&C^\alpha(\rho_{AB|C_1\cdots C_{N-2}})\\&&=(2T(\rho_{AB}))^{\frac{\alpha}{2}}\\
&&\geq|2T(\rho_A)-2T(\rho_{B})|^{\frac{\alpha}{2}}\\
&&=|C^2(\rho_{A|BC_1\cdots C_{N-2}})-C^2(\rho_{B|AC_1\cdots C_{N-2}})|^{\frac{\alpha}{2}}\\
&&\geq C^\alpha(\rho_{A|BC_1\cdots C_{N-2}})-C^\alpha(\rho_{B|AC_1\cdots C_{N-2}})\\
&&\geq h\sum_{i=1}^{k_1-1}C^\alpha(\rho_{AM_i})+C^\alpha(\rho_{AM_{k_1}})-C^\alpha(\rho_{B|AC_1\cdots C_{N-2}})\\
&&\geq h\sum_{i=1}^{k_1-1}C^\alpha(\rho_{AM_i})+C^\alpha(\rho_{AM_{k_1}})-J_B,
\end{eqnarray*}
where the first inequality is due to the left inequality in (\ref{LS}). From Lemma, one gets the second inequality. Using the inequality $(1+t)^x\geq 1+(2^{x}-1)t^x$, $t \geq 1$, $0\leq x \leq 1$, we get the third inequality. The last inequality is due to Theorem 1.

If $C(\rho_{A|BC_1\cdots C_{N-2}})\leq C(\rho_{B|AC_1\cdots C_{N-2}})$, similar to the above derivation, we can obtain another inequality in Theorem 2. $\Box$

Theorem 2 shows that the entanglement contained in the pure states $\rho_{ABC_1\cdots C_{N-2}}$ is related to the sum of entanglement between bipartitions of the system. The lower bound in inequalities (\ref{thm2}) is easily calculable. As an example, let us consider the four-qubit pure state $|\psi\rangle_{ABCD}=\frac{1}{\sqrt{2}}(|0000\rangle+|1001\rangle)$. We have $C(\rho_{AB})=C(\rho_{AC})=0$, $C(\rho_{AD})=1$, and $C_a(\rho_{BA})=C_a(\rho_{BC})=C_a(\rho_{BD})=0$. Therefore, $C(|\psi\rangle_{AB|CD})\geq 2^\frac{\alpha}{2}-1$, $0\leq\alpha\leq2$. Namely, the state $|\psi\rangle_{ABCD}$ saturates the inequality (\ref{thm2}) for $\alpha=2$.

Similar to the proof of Theorem 2, from (\ref{CAA}) we can derive another upper bound of the $\alpha$th power of concurrence as follows.

{\bf [Theorem 3]}. For any $N$-qubit state $|\psi\rangle_{ABC_1\cdots C_{N-2}}$, we have
\begin{eqnarray}\label{th}\nonumber
&&C^\alpha(|\psi\rangle_{AB|C_1\cdots C_{N-2}})\\ \nonumber
&&\geq\mathrm{max}\left\{\left(\sum_{i=1}^{N-2}C^2(\rho_{AC_i})+C^2(\rho_{AB})\right)^\frac{\alpha}{2}-J_B,\right.\\
&& \left.\left(\sum_{i=1}^{N-2}C^2(\rho_{BC_i})+C^2(\rho_{AB})\right)^\frac{\alpha}{2}-J_A\right\},
\end{eqnarray}
for $0\leq\alpha\leq2$, $N\geq 4$,
where $C_a^2(\rho_{AM_i})$ is defined in (\ref{coa}) and  $J_A= \sum_{i=1}^{k_1}h^{i-1}C^\alpha_a(\rho_{AM_i})$, $J_B=\sum_{i=1}^{k_2}h^{i-1}C^\alpha_a(\rho_{BM_i})$, $h=2^\frac{\alpha}{2}-1$.

{\it Example 2}. Let us consider the 4-qubit generalized $W$-class state,
\begin{eqnarray}\nonumber
|W\rangle_{ABC_1C_2}&=\lambda_1|1000\rangle+\lambda_2|0100\rangle\\
&+\lambda_3|0010\rangle+\lambda_4|0001\rangle,
\end{eqnarray}
where $\sum_{i=1}^4\lambda_i^2=1$. We have $C(|W\rangle_{AB|C_1C_2})=2\sqrt{(\lambda_1^2+\lambda_2^2)(\lambda_3^2+\lambda_4^2)}$, $C(\rho_{AB})=C_a(\rho_{AB})=2\lambda_1\lambda_2$, $C(\rho_{AC_1})=C_a(\rho_{AC_1})=2\lambda_1\lambda_3$, $C(\rho_{AC_2})=C_a(\rho_{AC_2})=2\lambda_1\lambda_4$. Taking $\lambda_1=\frac{3}{4},~\lambda_2=\frac{1}{2},~\lambda_3=\frac{\sqrt{2}}{4}$ and $\lambda_4=\frac{1}{4}$,
we get $J_A=J_B=\left(\frac{3}{8}\right)^\alpha+h\left(\frac{3\sqrt{2}}{8}\right)^\alpha+h^2\left(\frac{3}{4}\right)^\alpha$. 
Set $y_1=C^\alpha(|W\rangle_{AB|C_1C_2})-\left((C^2(\rho_{AB})+C^2(\rho_{AC_1})+C^2(\rho_{AC_2}))^\frac{\alpha}{2}-J_A\right)$ to be the difference between the left and right side of (\ref{th}). We have $y_1=\left(\frac{\sqrt{39}}{8}\right)^\alpha-\left(\frac{\sqrt{63}}{8}\right)^\alpha+\left(\frac{3}{4}\right)^\alpha+h\left(\frac{3\sqrt{2}}{8}\right)^\alpha+h^2\left(\frac{3}{8}\right)^\alpha$. From the inequality (9) in \cite{jin}, such difference is given by $y_2=\left(\frac{\sqrt{39}}{8}\right)^\alpha-\left(\frac{\sqrt{63}}{8}\right)^\alpha+\left(\frac{3}{8}\right)^\alpha
+\frac{\alpha}{2}\left(\frac{3\sqrt{2}}{8}\right)^\alpha+\left(\frac{\alpha}{2}\right)^2\left(\frac{3}{4}\right)^\alpha$. From Fig. 2 we can see that the difference between the left and right of the generalized monogamy inequality (\ref{th}) is smaller than that of the result from \cite{jin}.

\begin{figure}
  \centering
  \includegraphics[width=8.5cm]{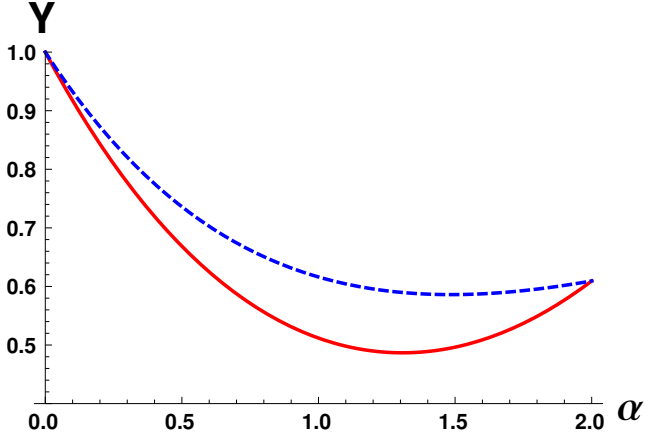}\\
  \caption{$Y$ stands for the differences between the left and right of the generalized monogamy inequalities:
  solid (red) line for (\ref{th}); dashed (blue) line for (9) in \cite{jin}.}\label{2}
\end{figure}

Theorem 2 and Theorem 3 present monogamy relations for $N$-qubit pure states under the partition $AB$ and $C_1...C_{N-2}$, which are different from the usual monogamy inequalities in Ref. \cite{JAB}.  Those results give rise to finer weighted characterizations of the entanglement distributions among the subsystems, as illustrated in Example 2. Moreover, Theorem 3 reduces to the result in \cite{zhu} as a special case of $\alpha=2$.
Theorem 2 and Theorem 3 give the monogamy-type lower bound of $C(|\psi\rangle_{AB|C_1\cdots C_{N-2}})$. According to the subadditivity of the linear entropy, we also have the following conclusion:

{\bf [Theorem 4]}. For any $2\otimes2\otimes\cdots\otimes2$ pure state $|\psi\rangle_{ABC_1\cdots C_{N-2}}$,  we have
\begin{eqnarray}\label{thm3}
C^\alpha(|\psi\rangle_{AB|C_1\cdots C_{N-2}})\leq J_A+J_B
\end{eqnarray}
for $0\leq\alpha\leq2$, $N\geq 4$, where $J_A$ and $J_B$ are defined similarly as in Theorem 2.

{\sf [Proof]}. Without loss of generality, there always exists a proper ordering of the subsystems such that $C_a^2(\rho_{AM_{t_1}})\geq \sum_{l={t_1}+1}^{k_1} C_a^2(\rho_{AM_l})$ and $C_a^2(\rho_{BM_{t_2}})\geq \sum_{l={t_2}+1}^{k_2} C_a^2(\rho_{BM_l})$ for any $1\leq t_1,t_2\leq k-1$ and $2\leq k_1,k_2\leq N-1$.
For qubit state $|\psi\rangle_{ABC_1\cdots C_{N-2}}$, one has
\begin{eqnarray*}
&&C^\alpha(|\psi\rangle_{AB|C_1\cdots C_{N-2}})\\&&=(2T(\rho_{AB}))^{\frac{\alpha}{2}}\\
&&\leq(2T(\rho_A)+2T(\rho_{B}))^{\frac{\alpha}{2}}\\
&&=(C^2(\rho_{A|BC_1\cdots C_{N-2}})+C^2(\rho_{B|AC_1\cdots C_{N-2}}))^{\frac{\alpha}{2}}\\
&&\leq C^\alpha(\rho_{A|BC_1\cdots C_{N-2}})+C^\alpha(\rho_{B|AC_1\cdots C_{N-2}})\\
&&\leq J_A+J_B,
\end{eqnarray*}
where the first inequality is due to the right inequality in (\ref{LS}). The second inequality is due to Lemma. Using the Theorem 1, one gets the last inequality. $\Box$

Let us consider the following four-qubit pure state, $|\psi\rangle_{ABCD}=\frac{1}{\sqrt{3}}(|0000\rangle+|0010\rangle+|1011\rangle)$. Then from the result in \cite{jin}, 
one gets $C^\alpha(|\psi\rangle_{AB|CD})\leq (\frac{2\sqrt{2}}{3})^\alpha+\frac{\alpha}{2}(\frac{2}{3})^\alpha$. While from our Theorem 4,
we have $C^\alpha(|\psi\rangle_{AB|CD})\leq (\frac{2\sqrt{2}}{3})^\alpha+h(\frac{2}{3})^\alpha$ for any $0\leq\alpha\leq 2$, where $h=2^{\frac{\alpha}{2}}-1$, see Fig. 3.

\begin{figure}
  \centering
  \includegraphics[width=7cm]{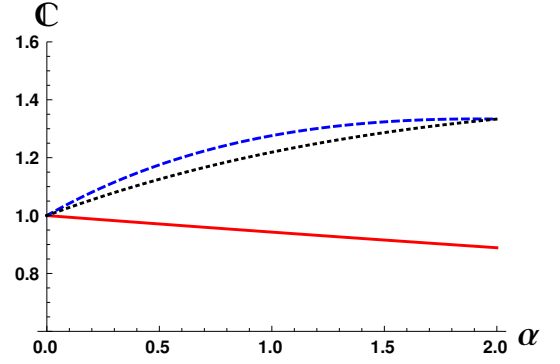}\\
  \caption{$C$ as a function of $\alpha$. Solid (Red) line is the value of $C^\alpha(|\psi\rangle_{AB|CD})$; Dotted (Black) line is the upper bound of (\ref{thm3}); Dashed (Blue) line is the upper bound of (11) in \cite{jin}
   .}\label{3}
\end{figure}

Now we generalize our results to the concurrence $C_{ABC_1|C_2\cdots C_{N-2}}(|\psi\rangle)$ under the partition $ABC_1$ and $C_2 \cdots C_{N-2}~(N\geq6)$ for pure state $|\psi\rangle_{ABC_1\cdots C_{N-2}}$.
Similar to Theorem 2, Theorem 3 and Theorem 4, we obtain the following corollaries:

{\bf [Corollary 1]}.  For any $N$-qubit pure state $|\psi\rangle_{ABC_1\cdots C_{N-2}}$, we have
\begin{eqnarray}\label{Co1}
&&C^\alpha(|\psi\rangle_{ABC_1|C_2\cdots C_{N-2}})\nonumber\\
&&\geq\mathrm{max}\left\{h\sum_{i=1}^{K_1-1}C^\alpha(\rho_{AM_i})+C^\alpha(\rho_{AM_{K-1}})-J_B,\right.\nonumber\\
&&\left. h\sum_{i=1}^{K_1-1}C^\alpha(\rho_{BM_i})+C^\alpha(\rho_{BM_{K-1}})-J_A\right\}-J_{C_1},
\end{eqnarray}
or
\begin{eqnarray}\label{CA1}
&&C^\alpha(|\psi\rangle_{ABC_1|C_2\cdots C_{N-2}})\nonumber\\
&&\geq\mathrm{max}\left\{\left(\sum_{i=1}^{N-2}C^2(\rho_{AC_i})+C^2(\rho_{AB})\right)^\frac{\alpha}{2}-J_B,\right.\nonumber\\
&&\left. \left(\sum_{i=1}^{N-2}C^2(\rho_{BC_i})+C^2(\rho_{AB})\right)^\frac{\alpha}{2}-J_A\right\}-J_{C_1},
\end{eqnarray}
where $J_A,~J_B$ are defined as in Theorem 2, $J_{C_1}=\sum_{i=1}^kh^{i-1}C^\alpha_a(\rho_{C_1M_i})$, $h=2^\frac{\alpha}{2}-1$, $2\leq m\leq N-3,~N\geq6$.

{\sf [Proof]}. For any $N$-qubit pure state $|\psi\rangle_{ABC_1\cdots C_{N-2}}$, if $C(|\psi\rangle_{AB|C_1\cdots C_{N-2}})\geq C(|\psi\rangle_{C_1|ABC_2\cdots C_{N-2}})$, we have
\begin{eqnarray*}
&&C^\alpha(|\psi\rangle_{ABC_1|C_2\cdots C_{N-2}})\\&&=(2T(\rho_{ABC_1}))^{\frac{\alpha}{2}}\\
&&\geq|2T(\rho_{AB})-2T(\rho_{C_1})|^{\frac{\alpha}{2}}\\
&&=|C^2(|\psi\rangle_{AB|C_1\cdots C_{N-2}})- C^2(|\psi\rangle_{C_1|ABC_2\cdots C_{N-2}})|^{\frac{\alpha}{2}}\\
&&\geq C^\alpha(|\psi\rangle_{AB|C_1\cdots C_{N-2}})-C^\alpha(|\psi\rangle_{C_1|ABC_2\cdots C_{N-2}}),
\end{eqnarray*}
where the first inequality is due to $T(\rho_{ABC_1})\geq T(\rho_{AB})-T(\rho_{C_1})$. Using Lemma, we get the second inquality. Combining Theorem 1 and Theorem 2, we obtain (\ref{Co1}), and combining Theorem 1 and Theorem 3, we obtain (\ref{CA1}). $\Box$

{\bf [Corollary 2]}. For any $N$-qubit pure state $|\psi\rangle_{ABC_1\cdots C_{N-2}}$, if $C(|\psi\rangle_{AB|C_1\cdots C_{N-2}})\leq C(|\psi\rangle_{C_1|ABC_2\cdots C_{N-2}})$, we have
\begin{eqnarray}\label{CA2}
&&C^\alpha(|\psi\rangle_{ABC_1|C_2\cdots C_{N-2}})\nonumber\\
&&\geq\left(C^2(\rho_{AC_1})+C^2(\rho_{BC_1})+\sum_{i=2}^{N-2}C^2(\rho_{C_1C_i})\right)^{\frac{\alpha}{2}}\nonumber\\
&&-J_A-J_B,
\end{eqnarray}
and
\begin{eqnarray}\label{CA3}
C^\alpha(|\psi\rangle_{ABC_1|C_2\cdots C_{N-2}})\leq J_A+J_B+J_{C_1},
\end{eqnarray}
where $J_A,~J_B$ are defined in Theorem 2, $J_{C_1}$ is defined in Corollary 1.

In Corollary 2, the upper bound is due to the right inequalities of (\ref{LS}) and (\ref{le}). Analogously, by using $T(\rho_{ABC_1})\geq |T(\rho_{AC_1})-T(\rho_B)|,~T(\rho_{ABC_1})\geq |T(\rho_{A})-T(\rho_{BC_1})|$, and $T(\rho_{ABC_1})\leq |T(\rho_{AC_1})+T(\rho_B)|,~T(\rho_{ABC_1})\leq |T(\rho_{A})+T(\rho_{BC_1})|$, one can get (\ref{CA2}) and (\ref{CA3}).

The lower bounds in Corollary 1 and Corollary 2 are not equivalent. We consider the following two examples to show that Corollary 1 and Corollary 2 give rise to different lower bounds.
Let us consider the pure state $|\psi\rangle_{ABC_1C_2C_3C_4}=\frac{1}{\sqrt{2}}(|000000\rangle+|101000\rangle)$. We have $C(|\psi\rangle)\geq1$ from (\ref{CA1}) and $C(|\psi\rangle)\geq0$ from (\ref{CA2}). Namely, bound (\ref{CA1}) is better than (\ref{CA2}) in this case. Nevertheless,
for the state $|\psi\rangle_{ABC_1C_2C_3C_4}=\frac{1}{\sqrt{2}}(|000000\rangle+|001100\rangle)$, one has $C(|\psi\rangle)\geq0$ from (\ref{CA1}) and $C(|\psi\rangle)\geq1$ from (\ref{CA2}). The bound (\ref{CA2}) is better than (\ref{CA1}) in this case.

\section{Tighter generalized monogamy and polygamy relations of negativity}

Another well-known quantifier of bipartite entanglement is the negativity. Given a bipartite state $\rho_{AB}$ in $H_A\otimes H_B$, the negativity is defined by \cite{GRF},
\begin{equation*}
 N(\rho_{AB})=\frac{||\rho_{AB}^{T_A}||-1}{2},
\end{equation*}
where $\rho_{AB}^{T_A}$ is the partially transposed matrix of $\rho_{AB}$ with respect to the subsystem $A$, $||X||$ denotes the trace norm of $X$, i.e $||X||=\mathrm{Tr}\sqrt{XX^\dag}$.
Negativity is a computable measure of entanglement, and is a convex function of $\rho_{AB}$. It vanishes if and only if $\rho_{AB}$ is separable for the $2\otimes2$ and $2\otimes3$ systems \cite{MPR}. For the purposes of discussion, we use the following definition of negativity:
$N(\rho_{AB})=||\rho_{AB}^{T_A}||-1$.

For any bipartite pure state $|\psi\rangle_{AB}$ in a $d\otimes d$ quantum system with Schmidt rank $d$,
$|\psi\rangle_{AB}=\sum_{i=1}^d\sqrt{\lambda_i}|ii\rangle$,
one has
\begin{eqnarray}\label{ne}
 N(|\psi\rangle_{AB})=2\sum_{i<j}\sqrt{\lambda_i\lambda_j},
\end{eqnarray}
from the definition of concurrence (\ref{CD}), we have
\begin{eqnarray}\label{co}
 C(|\psi\rangle_{AB})=2\sqrt{\sum_{i<j}\lambda_i\lambda_j}.
\end{eqnarray}
Combining (\ref{ne}) with (\ref{co}), one obtains
\begin{eqnarray}\label{nac}
N(|\psi\rangle_{AB})\geq C(|\psi\rangle_{AB}).
\end{eqnarray}

For any bipartite pure state    $|\psi\rangle_{AB}$ with Schmidt rank 2,
one has $N(|\psi\rangle_{AB})=C(|\psi\rangle_{AB})$ from (\ref{ne}) and (\ref{co}).
For a mixed state $\rho_{AB}$, the convex-roof extended negativity  (CREN) is defined by
\begin{equation*}
 N_c(\rho_{AB})=\mathrm{min}\sum_ip_iN(|\psi_i\rangle_{AB}),
\end{equation*}
where the minimum is taken over all possible pure state decompositions $\{p_i,~|\psi_i\rangle_{AB}\}$ of $\rho_{AB}$. CREN gives a perfect discrimination of positively partial transposed bound entangled states and separable states in any bipartite quantum systems \cite{PH,WJM}. For a mixed state $\rho_{AB}$, the convex-roof extended negativity of assistance (CRENOA) is defined by \cite{JAB}
\begin{equation*}
 N_a(\rho_{AB})=\mathrm{max}\sum_ip_iN(|\psi_i\rangle_{AB}),
\end{equation*}
where the maximum is taken over all possible pure state decompositions $\{p_i,~|\psi_i\rangle_{AB}\}$ of $\rho_{AB}$.

CREN is equivalent to concurrence for any pure state with Schmidt rank 2 \cite{JAB}. Consequently for any two-qubit mixed state $\rho_{AB}$, one has
\begin{eqnarray}\label{N1}
 N_c(\rho_{AB})=C(\rho_{AB})
\end{eqnarray}
and
\begin{eqnarray}\label{N2}
 N_a(\rho_{AB})= C_a(\rho_{AB}).
\end{eqnarray}

For $N$-qubit pure state $|\psi\rangle_{A|B_1B_2,\cdots,B_{N-1}}$, from (\ref{nac}), (\ref{N1}), (\ref{N2}) and the monogamy of the concurrence, we have
\begin{eqnarray}\label{negativity1}
&&N^\alpha(|\psi\rangle_{A|B_1B_2,\cdots,B_{N-1}})\nonumber\\&&\geq N_c^\alpha(\rho_{AB_1})+N_c^\alpha(\rho_{AB_2})+\cdots+N_c^\alpha(\rho_{AB_{N-1}}),
\end{eqnarray}
for $\alpha\geq2$.
The dual inequality \cite{JAB} in terms of CRENOA is given by
\begin{eqnarray}\label{negativity2}
&&N^2(|\psi\rangle_{A|B_1B_2,\cdots,B_{N-1}})\nonumber\\
&&\leq N_a^2(\rho_{AB_1})+N_a^2(\rho_{AB_2})+\cdots+N_a^2(\rho_{AB_{N-1}})\nonumber\\
&&=\sum_{i=1}^{k}N_a^2(\rho_{AM_i}),
\end{eqnarray}
where $N_a^2(\rho_{AM_i})=\sum_{j=M_{i-1}+1}^{M_i}N_a^2(\rho_{AB_j})$ with $M_0=0,~\sum_{i=1}^kM_i=N-1$, $1\leq k\leq N-1$. By similar consideration to concurrence, we get the upper bound of the $\alpha$th power of negativity as follows.

{\bf[Theorem 5]}. For any $N$-qubit pure state $|\psi\rangle_{AB_1B_2,\cdots,B_{N-1}}$, we have
\begin{eqnarray}\label{th4}
 && N^\alpha(|\psi\rangle_{A|B_1B_2,\cdots,B_{N-1}})\nonumber\\[1mm]
  &&\leq  N^\alpha_a(\rho_{AM_1})+hN_a^\alpha(\rho_{AM_2})+\cdots\nonumber\\[1mm]
  &&+h^{k-1}N_a^\alpha(\rho_{AM_k}),
\end{eqnarray}
for $0\leq \alpha\leq2$, where $h=2^\frac{\alpha}{2}-1$, $N\geq 4$.

{\bf [Theorem 6]}. For any qubit state $|\psi\rangle_{ABC_1\cdots C_{N-2}}$, we have
\begin{eqnarray}\label{thm6}\nonumber
&&N^\alpha(\rho_{AB|C_1\cdots C_{N-2}})\\ \nonumber
&&\geq\mathrm{max}\left\{h\sum_{i=1}^{k_1-1}N_c^\alpha(\rho_{AM_i})+N_c^\alpha(\rho_{AM_{k_1}})-J'_B,\right.\\
&& \left. h\sum_{i=1}^{k_2-1}N_c^\alpha(\rho_{BM_i})+N_c^\alpha(\rho_{BM_{k_2}})-J'_A\right\},
\end{eqnarray}
for $0\leq\alpha\leq2$, $N\geq 4$,
where  $J'_A= \sum_{i=1}^{k_1}h^{i-1}N^\alpha_a(\rho_{AM_i})$, $J'_B=\sum_{i=1}^{k_2}h^{i-1}N^\alpha_a(\rho_{BM_i})$, $h=2^\frac{\alpha}{2}-1$.

{\sf [Proof]}. Without loss of generality, there always exists a proper ordering of the subsystems 
such that $N_a^2(\rho_{AM_{t_1}})\geq \sum_{l={t_1}+1}^{k_1} N_a^2(\rho_{AM_l})$ and $N_a^2(\rho_{BM_{t_2}})\geq \sum_{l={t_2}+1}^{k_2} N_a^2(\rho_{BM_l})$, $1\leq t_i\leq k_i-1$  $(i=1,2)$, $2\leq k_1,k_2\leq N-1$.
For pure state $|\psi\rangle_{ABC_1\cdots C_{N-2}}$, we have
\begin{eqnarray*}
&&N^\alpha(|\psi\rangle_{AB|C_1\cdots C_{N-2}})\geq C^\alpha(|\psi\rangle_{AB|C_1\cdots C_{N-2}})\nonumber\\
&&\geq\mathrm{max}\left\{h\sum_{i=1}^{k_1-1}C^\alpha(\rho_{AM_i})+C^\alpha(\rho_{AM_{k_1}})-J_B,\right.\\
&& \left. h\sum_{i=1}^{k_2-1}C^\alpha(\rho_{BM_i})+C^\alpha(\rho_{BM_{k_2}})-J_A\right\}\nonumber\\
&&=\mathrm{max}\left\{h\sum_{i=1}^{k_1-1}N_c^\alpha(\rho_{AM_i})+N_c^\alpha(\rho_{AM_{k_1}})-J'_B,\right.\\
&& \left. h\sum_{i=1}^{k_2-1}N_c^\alpha(\rho_{BM_i})+N_c^\alpha(\rho_{BM_{k_2}})-J'_A\right\},
\end{eqnarray*}
where the first inequality is due to (\ref{nac}), the second inequality is from Theorem 2, the equality is based on (\ref{N1}) and  (\ref{N2}). $\Box$

{\bf [Theorem 7]}. For any qubit state $|\psi\rangle_{ABC_1\cdots C_{N-2}}$, we have
\begin{eqnarray*}
&&N^\alpha(|\psi\rangle_{AB|C_1\cdots C_{N-2}})\\ \nonumber
&&\geq\mathrm{max}\left\{\left(\sum_{i=1}^{N-2}N_c^2(\rho_{AC_i})+N_c^2(\rho_{AB})\right)^\frac{\alpha}{2}-J'_B,\right.\\
&& \left. \left (\sum_{i=1}^{N-2}N_c^2(\rho_{BC_i})+N_c^2(\rho_{AB})\right)^\frac{\alpha}{2}-J'_A\right\},
\end{eqnarray*}
for $0<\alpha\leq2$, $N\geq 4$,
where $J'_A= \sum_{i=1}^{k_1}h^{i-1}N^\alpha_a(\rho_{AM_i})$, $J'_B=\sum_{i=1}^{k_2}h^{i-1}N^\alpha_a(\rho_{BM_i})$, $h=2^\frac{\alpha}{2}-1$.

For $N$-qubit pure state $|\psi\rangle_{AB_1B_2,\cdots,B_{N-1}}$, based on the result in \cite{ce,zzj}, one has 
$N(|\psi\rangle_{AB|C_1\cdots C_{N-2}})\leq \sqrt{\frac{r(r-1)}{2}}C(|\psi\rangle_{AB|C_1\cdots C_{N-2}})$, where $r$ is the Schmidt rank of the pure state $|\psi\rangle_{ABC_1\cdots C_{N-2}}$. From Theorem 4, we can obtain the upper bound of negativity under the partition $AB$ and $C_1\cdots C_{N-2}$.

{\bf [Theorem 8]}. For any qubit state $|\psi\rangle_{ABC_1\cdots C_{N-2}}$, we have
\begin{eqnarray*}
N^\alpha(|\psi\rangle_{AB|C_1\cdots C_{N-2}})\leq \left(\frac{r(r-1)}{2}\right)^\frac{\alpha}{2}(J'_A+J'_B),
\end{eqnarray*}
for $0\leq\alpha\leq2$, where $J'_A,~J'_B$ are given in Theorem 6.

\section{conclusion}
Entanglement monogamy and polygamy relations are fundamental properties of multipartite entangled states. We have presented tighter monogamy relations of the $\alpha$th power of concurrence for $N$-qubit systems by showing the relations among $C(|\psi\rangle_{AB|C_1\cdots C_{N-2}}),~C(\rho_{AB}),~C(\rho_{AC_i}),~C(\rho_{BC_i}),~C_a(\rho_{AC_i})$, and $C_a(\rho_{BC_i})$, $1\leq i\leq N-2$, which give rise to the larger lower bounds and smaller upper bounds on the entanglement sharing among the partitions. The monogamy relations based on concurrence and COA have been investigated. We have obtained the smaller upper bound of the $\alpha$th ($0\leq\alpha\leq2$) power of concurrence based on COA. We then have derived the tighter monogamy and polygamy relations satisfied by the $\alpha$th ($0\leq\alpha\leq2$) power of concurrence in $N$-qubit pure states under the partition $AB$ and $C_1 \cdots C_{N-2}$, as well as under the partition $ABC_1$ and $C_1\cdots C_{N-2}$. These relations also give rise to a kind of trade-off relationship restricted by the lower and upper bounds of concurrences.
Based on the relations between negativity and concurrence, we have also obtained the similar results for CREN and CRENOA.
These results may be generalized to monogamy and polygamy relations under arbitrary partitions $C_{ABC_1\cdots C_i|C_{i+1}\cdots C_{N-2}},~2\leq i\leq N-2$.
Our approach may be also used for the investigation of entanglement distribution based on other measures of quantum correlations.

\bigskip
\noindent{\bf Acknowledgments}\, \, This work is supported by the NSF of China under Grant No. 11847209; 11675113; Beijing Municipal Commission of Education (KM201810011009) and the China Postdoctoral Science Foundation funded project.

\end{document}